\documentclass[sigconf,nonacm]{acmart}
\usepackage{cleveref}
\usepackage{algorithmic}
\usepackage{graphicx}
\usepackage{textcomp}
\usepackage{xcolor}
\usepackage{multirow}
\usepackage{enumitem}
\usepackage[htt]{hyphenat}
\usepackage{balance}
\usepackage{booktabs}
\usepackage{xspace}
\usepackage{tcolorbox}
\usepackage{subcaption}
\tcbuselibrary{skins, breakable}

\newcommand{\orgname}{\textsc{AutoFL}\xspace}
\newcommand{\name}{\textsc{AutoCrashFL}\xspace}
\newcommand{\baselineone}{\textsc{Baseline~1}\xspace}
\newcommand{\baselinetwo}{\textsc{Baseline~2}\xspace}

\newcommand{\company}{SAP HANA\xspace}

\AtBeginDocument{%
  }

\setcopyright{acmlicensed}
\copyrightyear{2018}
\acmYear{2018}
\acmDOI{XXXXXXX.XXXXXXX}
\acmConference[Conference acronym 'XX]{Make sure to enter the correct
  conference title from your rights confirmation email}{June 03--05,
  2018}{Woodstock, NY}
\acmISBN{978-1-4503-XXXX-X/2018/06}

\begin{document}

\title{Finding the Needle in the Crash Stack: Industrial-Scale Crash Root Cause Localization with \name}

\author{Sungmin Kang}
\email{sungmin@nus.edu.sg}
\orcid{0000-0002-0298-5320}
\affiliation{%
  \institution{NUS}
  \city{Singapore}
  \country{Singapore}
}

\author{Sumi Yun}
\email{sumi.yun@sap.com}
\author{Jingun Hong}
\email{jingun.hong@sap.com}
\affiliation{%
   \institution{SAP Labs Korea}
   \city{Seoul}
   \country{Republic of Korea}
}

\author{Shin Yoo}
\email{shin.yoo@kaist.ac.kr}
\orcid{0000-0002-0836-6993}
\affiliation{%
  \institution{KAIST}
  \city{Daejeon}
  \country{Republic of Korea}
}

\author{Gabin An}
\email{gabin_an@korea.ac.kr}
\orcid{0000-0002-6521-8858}
\authornote{Corresponding Author}
\affiliation{%
  \institution{Korea University}
  \city{Seoul}
  \country{Republic of Korea}
}

\renewcommand{\shortauthors}{Kang et al.}

\begin{abstract}
Fault Localization (FL) aims to identify root causes of program failures. FL typically targets failures observed from test executions, and as such, often involves dynamic analyses to improve accuracy, such as coverage profiling or mutation testing. 
However, for large industrial software, measuring coverage for every execution is prohibitively expensive, making the use of such techniques difficult.
To address these issues and apply FL in an industrial setting, this paper proposes \name, an LLM agent for the localization of crashes that only requires the crashdump from the Program Under Test (PUT) and access to the repository of the corresponding source code. We evaluate \name against real-world crashes of \company, an industrial software project consisting of more than 35 million lines of code. Experiments reveal that \name is more effective in localization, as it identified 30\% crashes at the top, compared to 17\% achieved by the baseline. Through thorough analysis, we find that \name has attractive practical properties: it is relatively more effective for complex bugs, and it can indicate confidence in its results. Overall, these results show the practicality of LLM agent deployment on an industrial scale.
\end{abstract}

\ccsdesc[500]{Software and its engineering~Maintaining software}
\ccsdesc[500]{Software and its engineering~Software testing and debugging}
\keywords{Fault Localization, Crash Debugging, Large Language Models, Automated Debugging, Industrial Software, Root Cause Analysis}


\maketitle

\section{Introduction}

Like all software systems, enterprise applications are susceptible to bugs and crashes that demand timely resolution. A crucial step in the debugging process is Fault Localization (FL)~\cite{wong2016survey}, i.e., identifying the code responsible for a failure. Empirical studies show developers spend a significant portion of their debugging time on this task~\cite{Bohme2017DbgBench}. To alleviate this burden, researchers have proposed a variety of automated FL techniques, including spectrum-based fault localization (SBFL)~\cite{Wong2014DStar, Jones2002Tarantula} and mutation-based fault localization (MBFL)~\cite{Moon2014MUSE, Papadakis2015Metallaxis}. 

More recently, LLM-based fault localization techniques have been proposed~\cite{kang2024quantitative, xu2025flexfl, qin2024agentfl}. These approaches demonstrate strong performance on open-source projects using failing test cases or bug reports. LLM-based FL offers greater flexibility than traditional FL methods: unlike SBFL, which requires expensive test coverage, or MBFL, which involves running numerous program variants, LLM-based approaches can operate directly on textual descriptions of failure symptoms. This enables integration with lightweight tools (e.g., file content retrieval) and makes them especially appealing in scenarios where runtime instrumentation is infeasible.

We set out to answer a key open question in this paper: to what extent do recent advances in LLM-based FL translate into practically usable tools that can help developers in real-world debugging scenarios? One such scenario is debugging crashes, which represents a common and high-priority failure mode in enterprise software systems. Particularly in industrial contexts, crashes often lack reproducing inputs due to client privacy or operational constraints. Instead, often the only available information is the \textit{crashdump}, which captures the program state at the time of failure. 
Even the coverage information of the crashed execution is often absent in production environments, due to the performance overhead of instrumentation. 
These limitations severely constrain conventional FL tools such as SBFL and MBFL, which require dynamic analyses and are inapplicable when such data is unavailable. However, they are rarely reflected in standard open-source benchmarks. This motivates our exploration of LLM-based FL in this more realistic and restricted setting.
In contrast to conventional tools, LLMs offer the potential to reason over static contextual information~\cite{kang2024quantitative, Xia2025Agentless}, such as stack traces, crash messages, and source code.

To this end, we evaluate the performance of the recent \orgname~\cite{kang2024quantitative} tool, an LLM-based FL technique, on crashdumps collected from \company~\cite{bach2022testing}, a large-scale commercial in-memory database system, consisting of more than 35 million lines of code, primarily implemented in C and C++. At a high level, \orgname operates by providing an LLM with contextual information about a bug and allowing the model to autonomously invoke \emph{functions} which retrieve additional evidence from the code repository and available coverage information, eventually synthesizing this information to identify the most likely location of the fault. We chose \orgname as the foundation for our LLM-based crash FL approach due to its lightweight design, high degree of customizability, and language-agnostic architecture, which make it especially suitable for integration into our industrial context.



Nevertheless, applying \orgname in our setting requires significant domain-specific adaptation. In its original form, \orgname takes error messages and stack traces as input, and uses coverage information from failing test cases to guide and constrain the fault search space. In contrast, our target industrial environment requires the use of extremely large crashdump files, which differ significantly in both scale and structure, while coverage data from the crash execution is not available. To bridge this gap, we replace the original function set used by \orgname with a new suite of tools specifically tailored for crash debugging. These include functions to access different segments of the given crashdumps, to retrieve relevant source code snippets from the repository, and to resolve symbol locations in the codebase to enable deep, context-aware source exploration. These modifications allow the LLM to reason more effectively over complex industrial crash data. For clarity, we hereafter refer to this adapted, crash-oriented variant of the tool as \name.

From the CI/CD system of \company, we collected 454 crashes that were reported and debugged between April 2023 and March 2024. Using these crashes, we evaluate \name against stack-trace-based heuristic baselines, and report that \name correctly ranked the root cause at the top for 134 crashes (30\%), outperforming the baselines, which achieved 49 (11\%) and 75 (17\%), respectively.
In summary, our contributions are as follows:
\begin{itemize}
    \item We identify a common industrial problem, FL for crashes, and detail how it differs from usual FL formulations;
    \item We propose \name, an FL technique specialized towards crashes, and which can handle an industrial scale;
    \item We perform a thorough evaluation of \name, identifying its efficacy and current limitations.
\end{itemize}

The remainder of this paper is organized as follows. Section~\ref{sec:background} provides academic background. Section~\ref{sec:methodology} presents the design of \name, and Section~\ref{sec:exp-setup} details the experimental setup. Section~\ref{sec:results} reports the evaluation results, and Section~\ref{sec:root_cause_explanation} analyzes the root-cause explanations. Finally, Section~\ref{sec:conclusion} concludes.

\section{Background}
\label{sec:background}
This section provides the academic background of this work.

\subsection{Fault Localization}
\label{sec:background_fl}
Fault Localization (FL) is part of the debugging process, in which a developer or automated process discerns which part of the software is responsible for undesirable behavior. 
As the developer study of B{\"o}hme et al.~\cite{Bohme2017DbgBench} demonstrates, FL takes a significant portion of time for developers. 
As such, automating FL has been studied since the Tarantula~\cite{Jones2002Tarantula} technique was introduced. 
Multiple different families of techniques have been developed by researchers, including spectrum-based fault localization (SBFL)~\cite{Jones2002Tarantula,Wong2014DStar}, mutation-based fault localization (MBFL)~\cite{Moon2014MUSE, Papadakis2015Metallaxis}, and learning based fault localization techniques~\cite{Xia2019DeepFL, Li2021rl4fl}. 
Relevant to this work, the strong performance of LLMs on many software engineering domains~\cite{fan2023large} has led to their utilization in automated FL as well~\cite{kang2024quantitative, qin2024agentfl}. In particular, this work adapts the AutoFL~\cite{kang2024quantitative} tool, described in greater detail in Section~\ref{sec:background_autofl}.

\subsection{LLM Agents for Automated Debugging}
\label{sec:background_llm}

The flexibility of LLMs has allowed their adoption for many software engineering tasks, such as bug reproduction~\cite{Kang2023aa,Kang2024aa}, fault localization~\cite{kang2024quantitative} and patch generation~\cite{Rondon2025wu,Bouzenia2024aa,Kang2025sn,Zhang2024aa}. While debugging requires inspection of a significant amount of information~\cite{boehme2017dbgbench}, LLMs have a limitation in that they can only process a limited amount of information at once. In response, LLM agents~\cite{feldt2023towardsagents} which overcome this issue have rapidly been developed. LLM agents use tools to autonomously retrieve information while performing the task. Starting with examples such as AutoSD~\cite{Kang2025sn} and ChatRepair~\cite{Xia2024ab}, the number of tools incorporated with LLM-based debugging techniques has increased~\cite{Li2025aa}. While there are many LLM agents, we are unaware of agents that take crashdumps and employ LLM agents to identify the buggy files, as is done in this work.

\subsection{AutoFL}
\label{sec:background_autofl}

\orgname~\cite{kang2024quantitative} is an FL technique that provides an LLM with tools that retrieve source code and documentation. This architectural decision aims to overcome a weakness of LLMs, in that they have a limited \emph{context length}, i.e., the amount of text they can process at once is limited. By allowing LLMs to invoke tools, \orgname allows the LLM to selectively retrieve important information without exceeding the context length.

In its original formulation, \orgname begins with two inputs: (i) a failing test and (ii) its corresponding error message and call stack. With this information, \orgname prompts the LLM to sequentially invoke tools and generate an explanation on how the bug occurs. The tools respectively show covered files, covered methods for a file, the source code of a particular method, and the documentation for a method (if it exists). Given the initial information, the LLM will iteratively use the tools to retrieve the relevant information for the bug, such as methods in the call stack. When the LLM judges that it has retrieved sufficient information to generate an explanation, it ceases to call a tool and generates a natural language description describing the mechanism of the bug. Upon this, the LLM is prompted to provide the precise signature of the buggy method. The resulting LLM response is treated as the predicted fault location, concluding a single `run' of \orgname. To improve performance and gauge confidence, multiple runs can be aggregated through a voting scheme. The aggregation through voting allows methods suggested over multiple runs to be suggested first, which is empirically correlated with performance.

In this work, we choose to modify \orgname for fault localization of crashes as it can be adapted to the scenario of \company with relative ease. First, \orgname already uses stack traces as part of its input, unlike other recent techniques~\cite{Xia2025Agentless, Zhang2024aa} which utilize developer-written bug reports. Second, its modular design allows easy adaptation for crashes: only the tools it uses need to be replaced, while the architecture can stay in place.

\subsection{Crash Analysis}
\label{sec:background_crash}


Meanwhile, program crashes have long been an issue for software~\cite{glerum2009crashes}, and as such, many techniques specialized in dealing with crashes have been proposed. First, crashdumps such as the ones used in this study are actually used by developers -- Schr{\"o}ter et al.~\cite{schroter2010stack} find that bug reports accompanied with crash stacks are resolved more often. As crashes are a particularly visible type of bug, there are many techniques that attempt to reproduce crashes as well. EvoCrash~\cite{Soltani2020EvoCrashTSE} is one such technique that analyzes crash stacks and uses EvoSuite as a backbone to generate crash-reproducing tests. Relevant to this work, there have been many techniques that use crash stacks in conjunction with other analyses to perform FL, but they all fail at the scale of \company's codebase and its specific debugging scenario. F3~\cite{jin2013f3} generates a crash-reproducing test, then performs SBFL and custom location filtering. In our industrial setting, tests are frequently unavailable and coverage is not collected, making such an approach infeasible. CrashLocator~\cite{wu2014crashlocator} performs static analysis and slicing to assign suspiciousness to functions, but without a test; we note that slicing at the scale of \company would be excessively time-consuming. The recent work of CrashTracker~\cite{yan2024crashtracker} similarly performs static analysis for FL, then uses LLMs to generate explanations. Unlike our work, the LLM is not directly involved in the FL process. Finally, Du et al.~\cite{du2023resolving} evaluate the performance of ChatGPT in resolving crash bugs gathered from Stack Overflow. Most notably, in these cases the crash-causing code is already largely localized at the function level, making the work inapplicable to our scenario.

\begin{figure*}[ht]
    \centering
    \includegraphics[width=0.9\linewidth]{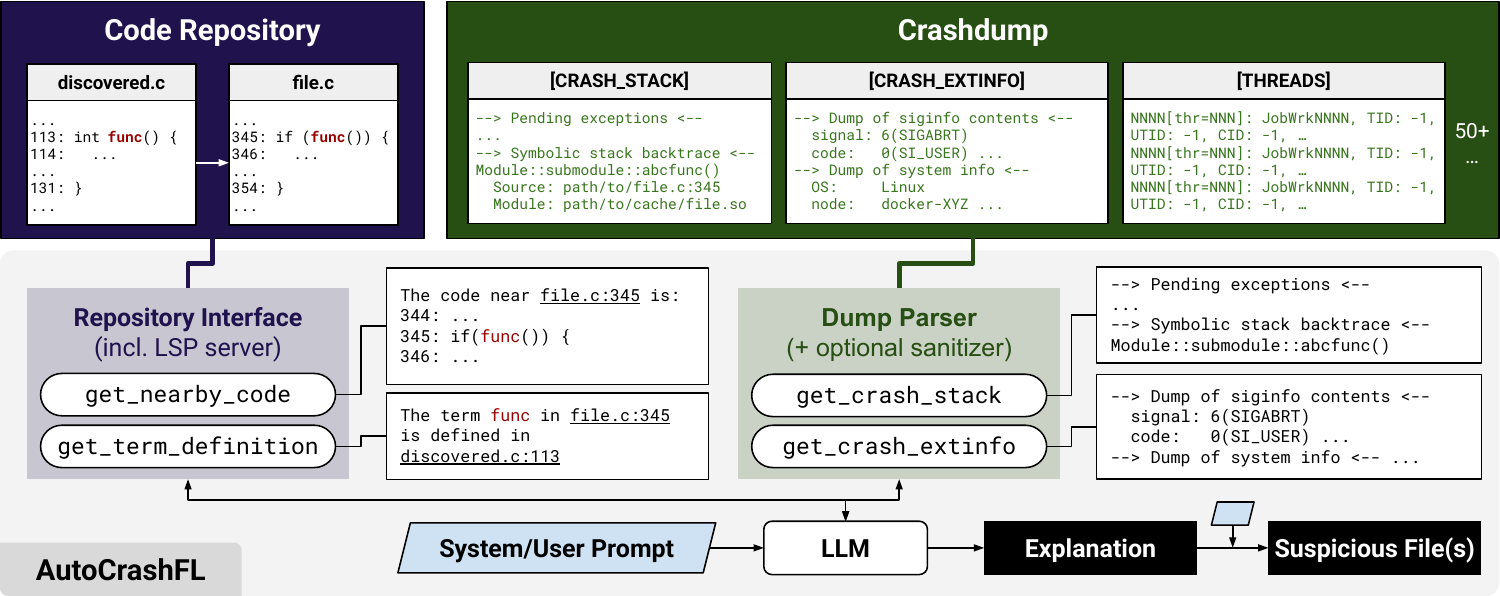}
    \caption{Diagram of \name.}
    \label{fig:diagram}
\end{figure*}

\section{Methodology}
\label{sec:methodology}

Figure~\ref{fig:diagram} shows the overall architecture of \name. We implement a parser for the crashdump, and a repository interface to access the code. The parser allows an LLM to easily access the information in the crashdumps by different segments -- similarly to \orgname, this is to mitigate the constraints from the context length. The repository interface allows an LLM to easily navigate the large code repository of \company. Based on these two toolkits, 
\name equips an LLM with four tools, allowing it to retrieve information and perform file-level fault localization. Finally, we run \name multiple times to generate independent reasoning trajectories, as prior work does~\cite{kang2024quantitative}, and aggregate the results into a final ranking of suspicious files. We describe each component of \name in more detail next.

\subsection{Crashdump Parser}
\label{sec:parser}

With \company, whenever an enterprise system crashes, the system information at the point of the crash is saved into a file known as a \emph{crashdump}. A crashdump is typically organized into more than 50 sections, including \texttt{BUILD}, \texttt{SYSTEMINFO}, \texttt{CRASH\_EXTINFO}, \texttt{CRASH\_STACK}, and \texttt{THREADS}, each providing different kinds of contextual information about the system state at the time of the crash. Professional developers look into these crashdumps to understand how the program crashed and how it can be fixed. 
As such, we opt to give the information in the crashdump to \name. However, crashdumps are often too long for an LLM to handle -- the median size of the crashdumps in our study is 6.6 megabytes, or about 3.1 million tokens, 24 times longer than the context length of \texttt{gpt-4o}. In fact, only one raw crashdump in our dataset fits in \texttt{gpt-4o}'s context length of 128,000. The longest crashdumps contain hundreds of megabytes worth of information, making them difficult to deal with, despite increases in LLM context lengths.

To mitigate this issue, we develop a crashdump parser, which is used to extract the most relevant information for FL. The parser is based on the observation that many sections of the crashdumps contain information that is only useful for debugging in special cases. For example, the \texttt{MOUNTINFO} section contains a substantial amount of information about the hardware configuration of the execution environment, and is consequently only useful for rare hardware bugs. To prevent confusion from such sections and include only the most relevant information, the parser first divides the crashdump into its constituent sections. Next, \name provides tools that retrieve the content of specific sections.



In particular, \name leverages two sections of the crashdump during operation: \texttt{CRASH\_EXTINFO} and \texttt{CRASH\_STACK}. Example excerpts of both sections are shown in \Cref{fig:diagram} within the Crashdump box.
The \texttt{CRASH\_EXTINFO} section provides background information about the crash, such as the process exit signal and system metadata (e.g., the accessed memory address in segmentation faults). This information enables the LLM to make informed decisions about the effect of the executed code.
The \texttt{CRASH\_STACK} section contains the symbolic stack backtrace at the point of failure, along with traces of unhandled pending exceptions. As \orgname~\cite{kang2024quantitative} begins FL from the failing test case, \name similarly requires concrete code locations from which \name can initiate its code search process using the repository interface. Among the available sections, we observe that \texttt{CRASH\_STACK} provides the richest such information. Across all crashes contained in our dataset, an average of 37\% of ground-truth buggy files had base filenames explicitly appearing in the stack trace, showing that while not exhaustive, stack traces expose a meaningful portion of buggy files. Prior crash fault-localization studies have likewise relied on stack traces as primary debugging cues~\cite{wu2014crashlocator}.

In our experiment, when providing the contents of \texttt{CRASH\_EXTINFO} and \texttt{CRASH\_STACK} to the LLM through tools, we considered two settings: raw section content (default) and sanitized content. The sanitized \texttt{CRASH\_STACK} retains only the source code locations (file path and line number) appearing in the stack trace while preserving their order, whereas the sanitized \texttt{CRASH\_EXTINFO} is reformatted into a simple dictionary structure without verbose headers such as \texttt{-{}-> Dump of siginfo contents <-{}-}.

While not directly used by \name, we note that the \texttt{BUILD} section, which contains the git hash for the code version, is used to set up the environment for our experimentation. Furthermore, the \texttt{CRASH\_EXTINFO} section is additionally used to distinguish bug types based on exit signal in Section~\ref{sec:exp-setup} and \ref{sec:results}.

\subsection{Repository Interface}

We provide two repository navigation tools to \name. The first tool takes as input a file path (within the target repository) and a line number, and retrieves the 10 preceding lines, the target line itself, and the 10 following lines. The second tool allows the LLM to point to a specific term within a line and use \texttt{clangd}, a compiler front-end for C/C++ implementing the Language Service Protocol (LSP),
to automatically identify which file and line a term is defined in. By providing these tools to the LLM, we designed \name such that the LLM could navigate the repository as necessary and reach a conclusion on fault localization.

Note that, unlike \orgname, we cannot rely on coverage information of the failing test, primarily due to the high overhead of coverage collection. \company spans more than 35 million LOC and 110 thousand files, occupying over 7GB. It is larger by orders of magnitude than commonly used open-source benchmarks: even the largest projects from the widely-used Defects4J~\cite{just2014defects4j},
BugsInPy~\cite{widyasari2020bugsinpy}, or SWE-bench~\cite{jimenez2024swebench} benchmarks have less than a million LOC. Measuring coverage for every execution run would therefore incur prohibitively high costs, and in practice, it is only performed on a weekly basis~\cite{bach2022testing}. With coverage unavailable, we instead provide additional mechanisms that allow navigation of deeper parts of the repository that are not directly exposed in the stack trace. While we considered RAG (Retrieval Augmented Generation~\cite{Lewis2020qy}) to provide code information, we decided against it due to the high cost of maintaining up-to-date indexing of the entire codebase, as well as the lack of control over the exact selection of code that is presented to the LLM. The \texttt{clangd}-based tool, on the other hand, fulfills this role by dynamically resolving symbol definitions and linking code references across the entire repository.


\subsection{\name}
Using the parsed information and the code navigation tools, \name prompts an LLM to identify the file that is responsible for the crash. In particular, the LLM is given the following tools:

\begin{itemize}[leftmargin=10pt]
    \item \texttt{get\_crash\_extinfo}: Returns either the raw (or sanitized) summary of the crash signal (e.g., signal number and code), along with host/system metadata (e.g., OS and kernel release/version) and processor information.
    \item \texttt{get\_crash\_stack}: Returns either the raw (or sanitized) stack report extracted from the crashdump, including symbolic backtraces of relevant threads at the point of failure and any pending exceptions.
    \item \texttt{get\_nearby\_code}: as a function for code navigation, this function takes as input a target file path and line number, and returns the nearby 10 lines of code above and below the input line.
    \item \texttt{get\_term\_definition}: as another function for code navigation, this function takes as input a target file path, line number, and the name of an identifier. Based on this, it will return the location in the repository in which the identifier was defined (if the language server, \texttt{clangd}, can find the location).
\end{itemize}

\begin{figure}[t]
    \includegraphics*[width=\linewidth]{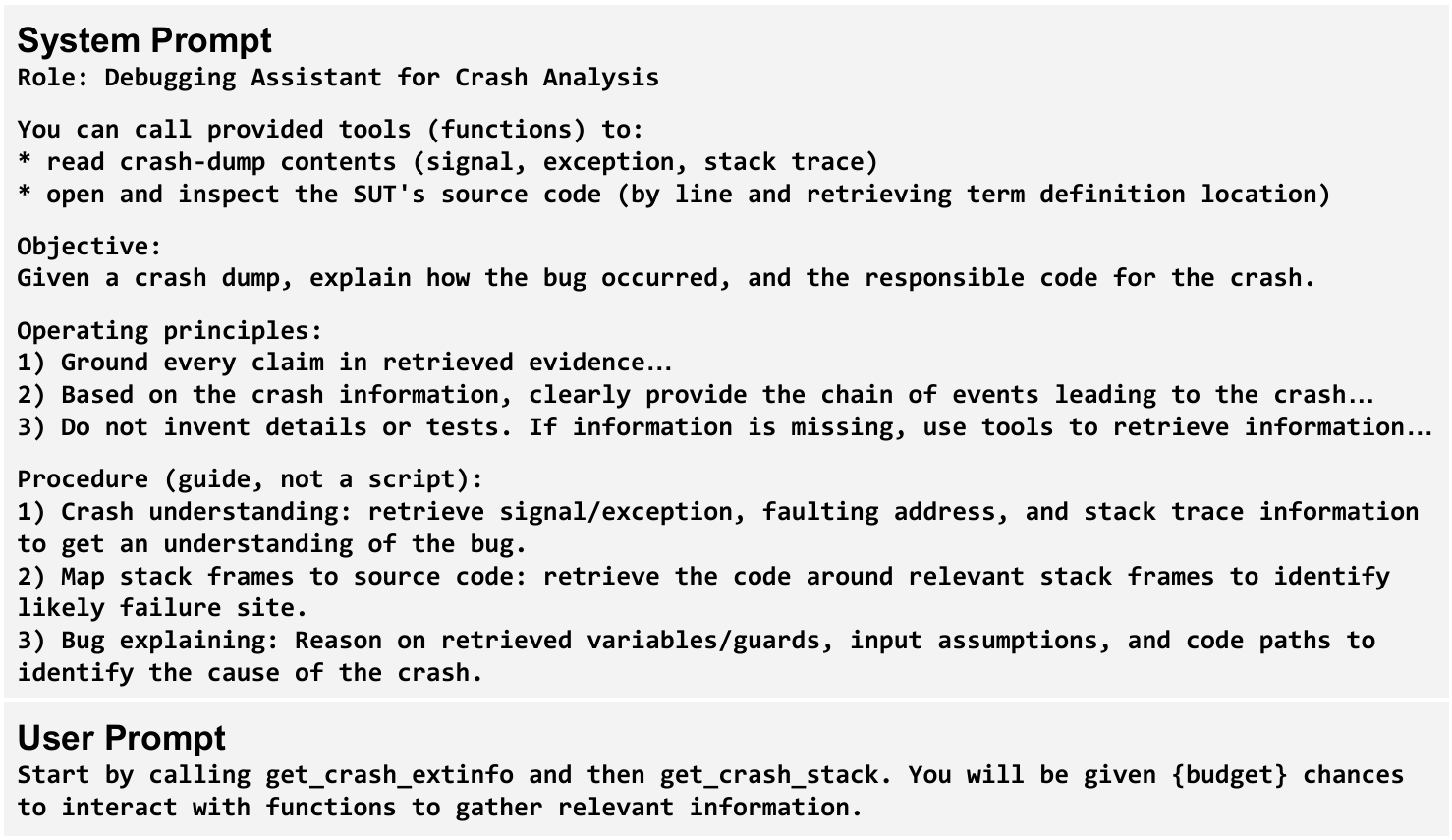}
    \caption{Prompts used by \name, lightly truncated for presentation.\label{fig:autocrashfl_prompts}}
  \end{figure}

Given this list of functions to call, the LLM autonomously retrieves information as it predicts is necessary. The prompts used to describe the role of the LLM are provided in Figure~\ref{fig:autocrashfl_prompts}. The system prompt states the general objective of the agent, while the user prompt provides instance-specific information such as the budget and instructs the LLM to start calling tools. Guided by these prompts, \name iteratively calls tools to gather information until it reaches a conclusion with an explanation of why the bug happened. As is the case for \orgname, after the explanation is generated, the LLM is prompted once again to suggest a list of culprit files (as paths) responsible for the bug.

\subsection{Ranking Aggregation}
\label{sec:ranking-aggregation}

LLM behavior can be stochastic, yielding different results in different runs. To stabilize and improve the performance of \name, given the same crash, we run the tool-calling process from start to finish $R$ times, and aggregate the results. In particular, for an LLM run that returns a set of suspicious files $S$, each file in the set is given a suspiciousness score of $1/|S|$; the scores for suggested files in all runs are added to give the final suspiciousness score for each file. Based on the score, we can derive two results. First, the score allows a ranking of all files that were suggested by the LLM in any run. Second, a `confidence score' can be calculated, defined as the suspiciousness of the file with the highest score divided by the number of runs, $R$. This confidence score is an indication of how consistently the LLM behaved. As we demonstrate in our results, confidence is correlated with better FL performance; thus, it can be used to present the highest-quality results to developers.

\begin{figure}[t]
  \includegraphics*[width=\linewidth]{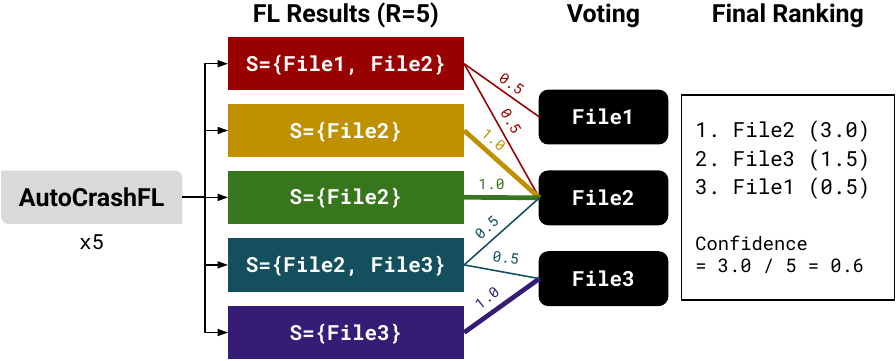}
  \caption{Rank Aggregation based on Voting\label{fig:rank_aggregation}}
\end{figure}

\section{Experimental Setting}
\label{sec:exp-setup}

We conduct an empirical evaluation of \name on the large C/C++ codebase \company to assess its feasibility for enterprise-scale crash debugging. This section provides details of the experimental setting.


\subsection{Research Questions}
\label{sec:rqs}

Our empirical evaluation is designed to answer the following research questions.

\begin{itemize}[leftmargin=10pt]
    \item \textbf{RQ1. Effectivenss of \name:} How effective is \name at localizing the root cause of collected crashes? We compare its localization accuracy to the heuristic baselines (described in \Cref{sec:baselines}).
    \item \textbf{RQ2. Performance Across Different Crash Types:} Are there specific types of crashes against which \name performs better or worse? We compare the performance of \name against different crash types and investigate the reason.
    \item \textbf{RQ3. Impact of Repeated Runs:} Do the repeated runs of \name and the voting-based aggregation contribute to its overall performance? To examine the cost–benefit trade-offs between repeated runs and accuracy, we compare the performance of \name under different numbers of runs ($R$).
    \item \textbf{RQ4. Contribution of Provided Functions:} Do all functions provided to \name contribute to its performance? We look at the patterns of tool usage, and present case studies of good and bad tool uses.
\end{itemize}

\subsection{Dataset Collection}

\begin{table}[t]
    \caption{Distribution of usable crash instances by crash type (Apr 2023–Mar 2024).}
    \centering
    \scalebox{0.93}{
    \begin{tabular}{l r r}
    \toprule
    Crash Type & Count & \% \\
    \midrule
    SIGABRT            & 289 & 63.7 \\
    SIGSEGV (NPE)      & 118 & 26.0 \\
    SIGSEGV (Non-NPE)  & 43  & 9.5  \\
    SIGBUS             & 3   & 0.7  \\
    SIGFPE             & 1   & 0.2  \\
    \midrule
    \textbf{Total}     & 454 & 100 \\
    \bottomrule
    \end{tabular}
    }
    \label{tab:crash-dataset}
\end{table}

We collect actual \company crash reports that had been previously debugged by developers and thus could be linked to their corresponding fixing commits. In total, we investigate 527 crash instances reported during the one-year period from April 2023 to March 2024, which represents a recent yet sufficiently mature time window at the time of initiating our study (i.e., enough time had passed for fixes to be produced by developers). After excluding 56 instances with invalid Git hashes in the crash dumps (e.g., dummy or temporary commit identifiers not present in the remote repository) and 17 instances for which the language server fails to initialize (mainly due to the old \texttt{clang} version), 454 crashes remain usable for evaluation. On average, these usable crash instances require changes to 4.37 files in their corresponding fixing commits.

We categorize the crash instances by crash type and present their distribution in Table~\ref{tab:crash-dataset}. The most frequent categories are SIGABRT (abort signal) and SIGSEGV (segmentation fault), whereas SIGBUS (bus error) and SIGFPE (computational error) are rarely observed. Through interviews with developers who regularly debug such crashes, we learn that SIGSEGV crashes are typically further classified based on the \textit{faulting address}, which is usually the first detail examined during debugging. If the recorded address is either \texttt{0x00} (Null) or a close-to-zero value (e.g., \texttt{0x3d}, representing \texttt{Null+offset}), the crash is identified as a \textit{null-pointer dereference}, aka a Null-Pointer Exception (NPE). Otherwise, it is treated as a non-NPE segmentation fault. Following this practice, we subdivide the collected SIGSEGV crashes into NPE and non-NPE categories.

\subsection{\name Parameters}

We evaluate \name with the maximum number of function interactions set to $N = 25$, a setting that ensures sufficient exploration of the code context without incurring excessive inference costs. Each crash is analyzed over $R = 10$ repeated runs, allowing us to aggregate rankings across stochastic LLM outputs and obtain more stable results, as described in Section~\ref{sec:ranking-aggregation}. By default, the functions \texttt{get\_crash\_extinfo} and \texttt{get\_crash\_stack} return the \textbf{raw} contents of their respective sections. For all experiments, we use OpenAI's \texttt{gpt-4o} as the underlying LLM.  

For RQ3, we investigate the effect of repeated runs by varying $R$ from 1 to 10 and comparing the FL accuracy of the resulting aggregated rankings. For RQ4, we keep all other settings identical to the default configuration, but configure \texttt{get\_crash\_extinfo} and \texttt{get\_crash\_stack} to return sanitized contents of their corresponding crashdump sections. In addition, we disable the deep-search functionality (i.e., we do not provide the \texttt{get\_term\_definition} tool to the LLM).

\subsection{Evaluation Metric}
We use the widely adopted FL accuracy metric, \texttt{acc@$k$} (or top-$k$ accuracy), here defined as the number of crash instances for which at least one actual faulty file (contained in the corresponding fixing commit) appears within the top-$k$ ranked results. We evaluate \texttt{acc@$k$} for $k \in \{1, 2, 3, 5, 10\}$. We also report the corresponding percentage of the total number of evaluated crash instances, for easier comparison of FL performance between different crash types.

\subsection{Heuristic Baselines}
\label{sec:baselines}

\begin{figure}
    \centering
    \includegraphics[width=0.8\linewidth]{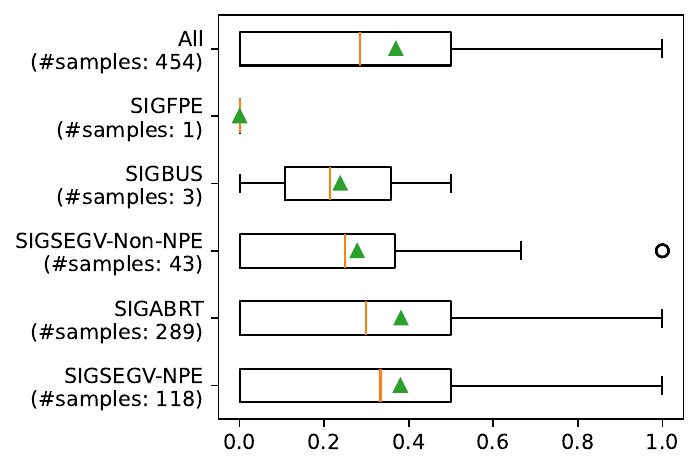}
    \caption{Ratio of buggy files whose base filenames explicitly appear in the \texttt{CRASH\_STACK}. The green triangle denotes the mean value, while the orange line indicates the median.}
    \label{fig:in_trace_ratio}
\end{figure}

As discussed in \Cref{sec:parser}, the crash stack in a crashdump file often provides useful hints about the location of the buggy file. In our scenario, a typical stack trace, provided by the \texttt{CRASH\_STACK} section of the crashdump, contains two main components: (1) a \textit{symbolic stack backtrace}, representing the call stack at the point where the crash occurs, and (2) a list of \textit{pending exceptions}, which captures the stack trace of unhandled exceptions that were pending at the time of the crash. \Cref{fig:in_trace_ratio} shows the ratio of buggy files whose base filenames explicitly appear in the \texttt{CRASH\_STACK}, both overall and by crash type. 
The value of \name lies in providing additional benefit beyond simply investigating the files appearing in the stack trace. Therefore, to more rigorously assess its accuracy, we compare \name against the following baselines, designed through consultation with \company developers.



\begin{itemize}[leftmargin=10pt]
    \item \textbf{\baselineone: Symbolic Stack Backtrace Only.} This baseline prioritizes files based solely on their position in the symbolic stack backtrace. Files are ranked in the order they appear in the stack trace, from top (most recent call) to bottom. This follows the observation of Wu et al.~\cite{wu2014crashlocator}, who note that functions closer to the crash point are more likely to be faulty.  Duplicate files are removed while preserving order. The resulting list serves as the FL ranking.

    \item \textbf{\baselinetwo: Pending Exceptions First, Symbolic Stack Backtrace Second.} In addition to the top-to-bottom assumption, \baselinetwo further utilizes files that appear in the pending exceptions stack trace. It assumes they are more indicative of the root cause as they reflect the initial deviation from expected behavior. Accordingly, we first extract files from the pending exceptions trace, followed by those from the symbolic stack backtrace, mapping each to its source file and removing duplicates. The concatenated list is used as the FL ranking.
\end{itemize}

\section{Results}
\label{sec:results}
This section provides the results for each research question.
\subsection{RQ1: Effectiveness of \name}

\begin{table}[t]
    \centering
    \caption{Comparison of \name and heuristic baselines in terms of localization accuracy, reported as \texttt{acc@$k$} (\texttt{a@k}). Results are based on 454 crash instances. The second column (Len.) denotes the average length of FL rankings produced by each technique.}
    \scalebox{0.93}{
    \begin{tabular}{c|r|rrrrr}
    \toprule
    Technique & Len. & \texttt{a@1} & \texttt{a@2} & \texttt{a@3} & \texttt{a@5} & \texttt{a@10}\\
    \midrule
    \multirow{2}{*}{\baselineone} & \multirow{2}{*}{14.76} & 49 & 88 & 142 & 192 & 255 \\
                               & & (11\%) & (19\%) & (31\%) & (42\%) & (56\%) \\\midrule
    \multirow{2}{*}{\baselinetwo} & \multirow{2}{*}{16.49} & 75 & 138 & 196 & 243 & \textbf{295} \\
                               & & (17\%) & (30\%) & (43\%) & (54\%) & (65\%) \\
                               \midrule
    \multirow{2}{*}{\name}     & \multirow{2}{*}{3.83}  & \textbf{134} & \textbf{195} & \textbf{227} & \textbf{245} & 255 \\
                               & & (30\%) & (43\%) & (50\%) & (54\%) & (56\%)\\
    \bottomrule
    \end{tabular}
    }
    \label{tab:RQ1_results}
\end{table}

\Cref{tab:RQ1_results} shows the performance of \name compared to the heuristic baselines. \name clearly outperforms both baselines in terms of \texttt{acc@k} for $k \in \{1, 2, 3, 5\}$. In particular, the strong \texttt{acc@1} demonstrates that \name can pinpoint the buggy file even when it does not appear at the top of either the symbolic backtrace or the pending-exception list.
Between the two baselines, \baselinetwo consistently outperforms \baselineone, suggesting that prioritizing files appearing in the pending exceptions is more effective in practice than considering only the files in the symbolic stack at the crash point.

\begin{figure}[t]
    \centering
    \includegraphics[width=0.5\linewidth]{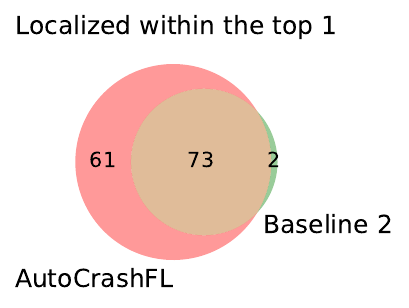}
    \caption{Venn diagram of crash instances localized at the top by \name and \baselinetwo}
    \Description{Venn diagram of crash instances localized at the top by \name and \baselinetwo}
    \label{fig:RQ1_venn}
\end{figure}

\Cref{fig:RQ1_venn} further analyzes the overlap between \name and \baselinetwo for top-1 localization. Out of 454 crashes, \name uniquely localizes 61 crashes at the top rank, while \baselinetwo uniquely localizes 2 crashes. The two methods agree on 73 crashes, indicating that \name covers most of \baselinetwo's successes while also expanding coverage to additional crashes that the heuristic fails to capture. This demonstrates that \name not only achieves higher accuracy overall, but also contributes unique localizations beyond what stack-based heuristics can provide.

\begin{table}[t]
    \centering
    \caption{Performance of augmented \name rankings. “Aug.” denotes that the \name ranking is extended with the files from the corresponding baseline’s ranking.}
    \scalebox{0.93}{
    \begin{tabular}{c|r|rrrrr}
    \toprule
    Technique & Len. & \texttt{a@1} & \texttt{a@2} & \texttt{a@3} & \texttt{a@5} & \texttt{a@10}\\
    \midrule
    \multirow{2}{*}{\name}     & \multirow{2}{*}{3.83}  & \textbf{134} & 195 & 227 & 245 & 255 \\
                               & & (30\%) & (43\%) & (50\%) & (54\%) & (56\%)\\\midrule
    \name & \multirow{2}{*}{15.96} & \textbf{134} & \textbf{198} & 234 & 258 & 292 \\
     Aug. w/ \baselineone    &     & (30\%) & (44\%) & (52\%) & (57\%) & (64\%) \\\midrule
    \name & \multirow{2}{*}{16.49} & \textbf{134} & \textbf{198} & \textbf{235} & \textbf{263} & \textbf{308} \\
     Aug. w/ \baselinetwo   &      & (30\%) & (44\%) & (52\%) & (58\%) & (68\%)\\
    \bottomrule
    \end{tabular}
    }
    \label{tab:RQ1_aug_results}
\end{table}

The only exception is \texttt{acc@10}, where \baselinetwo achieves better performance. This can be explained by the fact that \name produces rankings that are, on average, less than one-third the length of those generated by the baselines (3.83 vs. 14–16 files). While it is notable that \name improves accuracy while simultaneously reducing developer effort by providing shorter rankings, this conciseness may also omit valuable contextual information available in the original stack trace.

In the original \orgname study, this limitation was mitigated by augmenting the generated rankings with additional candidates that were not directly identified by \orgname but were covered by the failing test. Following this strategy, we also augment the \name-generated rankings with the files included in each baseline. For example, if the stack trace contains five files $f_1$, $f_2$, $f_3$, $f_4$, $f_5$, in order from top to bottom, and \name produces the ranking $[f_1, f_5]$, the remaining files are appended to the rankings, resulting in the augmented ranking $[f_1, f_5, f_2, f_3, f_4]$.

\Cref{tab:RQ1_aug_results} reports the performance of \name when its rankings are augmented with the file rankings from each baseline. When augmented, the \texttt{acc@1} remains unchanged at 134, but top-2 and top-3 improve slightly. At larger $k$ values, the augmented versions yield substantial gains. In particular, when \name is augmented with \baselinetwo, \texttt{acc@10} reaches 308 (68\%), the best among all settings. Overall, these results confirm that augmentation helps recover the contextual coverage that may be lost in \name's concise rankings, especially at larger $k$. While the augmented versions surpass the baselines, they do so by increasing the ranking length and thus the developer's inspection cost. In practice, this highlights a trade-off: unaugmented \name offers concise, high-precision rankings (lower effort), while augmented versions achieve higher recall at the cost of longer candidate lists. \textbf{In the remaining sections, we focus only on the original \name-generated rankings to clearly demonstrate its performance in isolated settings.}

\subsection{RQ2: Performance Across Crash Types}
\begin{table*}[t]
    \centering
    \caption{Performance of \name compared to heuristic baselines across different crash types, reported is \texttt{acc@$k$} (\texttt{a@k}). Each block corresponds to one crash type: SIGABRT, SIGSEGV due to Null Pointer Exceptions (NPE), SIGSEGV due to other errors (non-NPE), and two minor types (SIGBUS/SIGFPE). The column \texttt{Len.} denotes the average length of FL rankings produced by each technique, and numbers in parentheses indicate percentages relative to the total number of crashes in each category.}
    \scalebox{0.9}{
        \begin{tabular}{c|r|rrrrr|r|rrrrr}
        \toprule
         \multirow{2}{*}{Technique} 
         & \multicolumn{6}{c|}{\textbf{SIGABRT (total = 289)}} 
         & \multicolumn{6}{c}{\textbf{SIGSEGV (NPE) (total = 118)}} \\
         \cmidrule{2-13}
         & Len. & \texttt{a@1} & \texttt{a@2} & \texttt{a@3} & \texttt{a@5} & \texttt{a@10} 
         & Len. & \texttt{a@1} & \texttt{a@2} & \texttt{a@3} & \texttt{a@5} & \texttt{a@10} \\
        \midrule
        \multirow{2}{*}{\baselineone} 
          & \multirow{2}{*}{14.58} & 6 & 11 & 53 & 92 & 145 
          & \multirow{2}{*}{14.41} & 33 & \textbf{61} & 70 & \textbf{80} & \textbf{83} \\
          & & (2\%) & (4\%) & (18\%) & (32\%) & (50\%) 
          & & (28\%) & (52\%) & (59\%) & (68\%) & (70\%) \\\midrule
        \multirow{2}{*}{\baselinetwo} 
          & \multirow{2}{*}{17.01} & 33 & 64 & 108 & 143 & \textbf{184} 
          & \multirow{2}{*}{14.91} & 32 & 58 & 69 & 80 & 84 \\
          & & (11\%) & (22\%) & (37\%) & (49\%) & (64\%) 
          & & (27\%) & (49\%) & (58\%) & (68\%) & (71\%) \\\midrule
        \multirow{2}{*}{\name} 
          & \multirow{2}{*}{4.13} & \textbf{82} & \textbf{115} & \textbf{135} & \textbf{148} & 155 
          & \multirow{2}{*}{3.01} & \textbf{38} & \textbf{61} & \textbf{72} & 74 & 75 \\
          & & (28\%) & (40\%) & (47\%) & (51\%) & (54\%) 
          & & (32\%) & (52\%) & (61\%) & (63\%) & (64\%) \\\midrule\midrule
         \multirow{2}{*}{Technique} 
         & \multicolumn{6}{c|}{\textbf{SIGSEGV (Non-NPE) (total = 43)}}
         & \multicolumn{6}{c}{\textbf{SIGBUS/SIGFPE (total = 4)}}\\
         \cmidrule{2-13}
         & Len. & \texttt{a@1} & \texttt{a@2} & \texttt{a@3} & \texttt{a@5} & \texttt{a@10} 
         & Len. & \texttt{a@1} & \texttt{a@2} & \texttt{a@3} & \texttt{a@5} & \texttt{a@10} \\\midrule
        \multirow{2}{*}{\baselineone} 
          & \multirow{2}{*}{17.00} & 9 & 15 & 18 & 19 & \textbf{25} 
          & \multirow{2}{*}{13.50} & \textbf{1} & \textbf{1} & \textbf{1} & 1 & \textbf{2} \\
          & & (21\%) & (35\%) & (42\%) & (44\%) & (58\%) 
          & & (25\%) & (25\%) & (25\%) & (25\%) & (50\%) \\\midrule
        \multirow{2}{*}{\baselinetwo} 
          & \multirow{2}{*}{17.60} & 9 & 15 & 18 & 19 & \textbf{25}
          & \multirow{2}{*}{13.50} & \textbf{1} & \textbf{1} & \textbf{1} & 1 & \textbf{2} \\
          & & (21\%) & (35\%) & (42\%) & (44\%) & (58\%) 
          & & (25\%) & (25\%) & (25\%) & (25\%) & (50\%) \\\midrule
        \multirow{2}{*}{\name} 
          & \multirow{2}{*}{3.90} & \textbf{13} & \textbf{18} & \textbf{19} & \textbf{21} & 23
          & \multirow{2}{*}{3.25} & \textbf{1} & \textbf{1} & \textbf{1} & \textbf{2} & \textbf{2}\\
          & & (31\%) & (43\%) & (45\%) & (50\%) & (55\%) 
          & & (25\%) & (25\%) & (25\%) & (50\%) & (50\%) \\
        \bottomrule
        \end{tabular}
    }
    \label{tab:RQ2_results}
\end{table*}

\Cref{tab:RQ2_results} shows the performance of \name compared to the heuristic baselines across different crash types. Since the sample sizes of SIGBUS and SIGFPE were too small, we report them together. Overall, \name outperforms both baselines in terms of \texttt{acc@k} for smaller values of $k$, indicating that \name more accurately pinpoints the actual faulty location. As shown in RQ1, performance for larger $k$ can be further improved by augmenting the ranking list with baseline results; however, we do not consider such augmentation here.

Notably, \name's relative benefit is most prominent for SIGABRT crashes. For example, \name achieves \texttt{acc@1} of 82 cases (28\%) on SIGABRT, compared to only 6 (2\%) and 33 (11\%) for \baselineone and \baselinetwo, respectively. Between the two baselines, \baselinetwo significantly outperforms \baselineone, suggesting that for SIGABRT crashes, the true buggy location is more often contained in the pending-exception list rather than the symbolic backtrace.
This is because SIGABRT is frequently triggered by explicit calls to \texttt{abort()} 
or by runtime checks (e.g., failed assertions), where the symbolic backtrace 
only shows the abort site, while the pending exceptions preserve the 
context of the original faulty operation.

For SIGSEGV (NPE) crashes, the performance gap narrows. While \baselineone and \baselinetwo achieve competitive results at larger $k$ values (as also shown in RQ1), \name still provides the best top-1 accuracy (38 cases, 32\%) and maintains steady improvements through \texttt{acc@10}. Interestingly, in this category \baselineone slightly outperforms \baselinetwo. This is because when a program dereferences a null pointer, the segmentation fault occurs immediately, and the symbolic backtrace typically records the precise function and location of the invalid access. Indeed, for SIGSEGV (NPE), many crashes had no pending exceptions at all, as reflected by the relatively small difference in the ranking lengths produced by the two baselines compared to other crash types. Overall, SIGSEGV (NPE) represents the easiest category for fault localization, both for the baselines and for \name.

For SIGSEGV (non-NPE) crashes, baseline methods perform worse than for SIGSEGV (NPE) despite both being segmentation faults, likely due to the inherent difficulty of this category. As shown in \Cref{fig:in_trace_ratio}, the average ratio of ground-truth buggy files whose base filenames explicitly appear in the stack trace is relatively low, even though more files tend to be included in the trace (as reflected by the longer FL rankings of the baselines). A plausible explanation is that while these crashes are also segmentation faults, the faulting address is often a large arbitrary value, making its connection to the intended memory access ambiguous. In such cases (e.g., buffer overflows, dangling pointers, or incorrect offset calculations), the top frame in the crash stack may only indicate where the invalid memory access was detected, whereas the true defect lies earlier in the execution path. This ambiguity makes SIGSEGV (non-NPE) crashes more difficult to localize than SIGSEGV (NPE) crashes when relying solely on the crash stack, thereby explaining the baselines' lower accuracy. Nevertheless, \name substantially improves top-1 accuracy to 31\%, reaching a level comparable to NPE bugs and highlighting its value.
Finally, for SIGBUS and SIGFPE crashes, only four samples were available, making them rare crash types and limiting the ability to draw generalized conclusions from the observed performance. For these crashes, \name performed on par with or better than the baselines.

Taken together, these results suggest that \name generalizes well across different crash types, with particularly large advantages for SIGABRT and SIGSEGV (non-NPE) cases, while still providing competitive performance for SIGSEGV (NPE).

\subsection{RQ3: Impact of Repeated Runs}
\begin{figure}[t]
    \centering
    \includegraphics[width=1.0\linewidth]{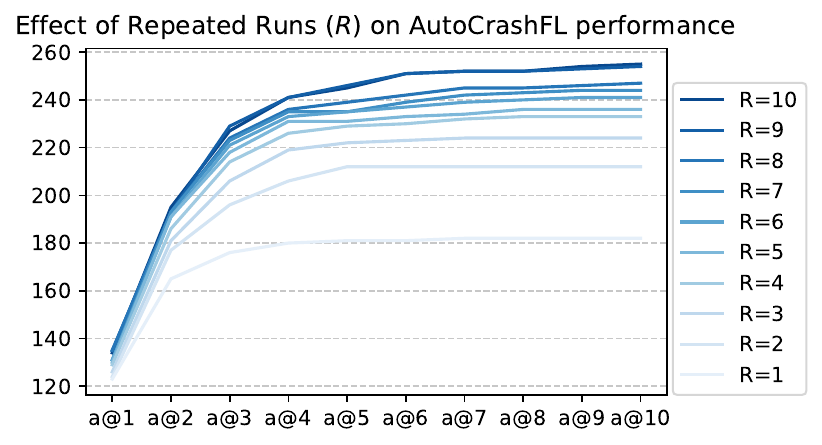}
    \caption{Effect of $R$ on \name's FL performance}
    \Description{Effect of $R$ on \name's FL performance}
    \label{fig:RQ3_rerun}
\end{figure}

In its original paper, \orgname reported improved performance when the number of repeated runs $R$ used for aggregation was increased~\cite{kang2024quantitative}. We investigate whether \name exhibits a similar trend; specifically, whether aggregating more runs for each crash improves FL accuracy.
\Cref{fig:RQ3_rerun} shows that FL accuracy consistently improves as $R$ increases. For example, \texttt{acc@1} rises from 123 crashes with $R=1$ to 134 crashes with $R=10$, and acc@10 increases from 182 to 255. These results confirm that aggregating multiple runs 
yields more stable and accurate rankings. 
Because the computational cost increases approximately linearly with $R$, it is advisable to perform as many runs as the available budget allows and apply voting-based aggregation to improve reliability. However, the marginal gains diminish as $R$ grows, aligning with the diminishing returns observed in the original study.

\begin{figure}[t]
    \includegraphics[width=0.9\linewidth]{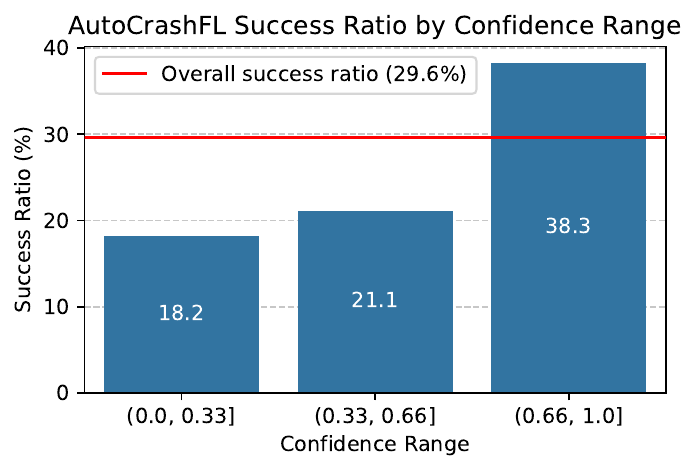}
    \caption{Success ratio of \name across different confidence ranges. Success is defined as ranking the actual buggy file at the top position.}
    \Description{Success ratio of \name across different confidence ranges. Success is defined as ranking the actual buggy file at the top position.}
    \label{fig:RQ3_confidence}
\end{figure}

As described in \Cref{sec:ranking-aggregation}, after aggregation, we also measure \name's confidence, defined as the maximum suspiciousness score among its FL outputs, following the strategy of \orgname. To assess whether this confidence score correlates with actual performance, we partition the score range into three buckets: $(0, 0.33]$ (\textit{low}), $(0.33, 0.66]$ (\textit{medium}), and $(0.66, 1.00]$ (\textit{high}), ensuring that each bucket contains at least 30 samples. \Cref{fig:RQ3_confidence} shows the success rate for each bucket, where success means that \name ranks the correct buggy file at the top. While the overall success rate is 29.6\%, it rises to 38.3\% in the \textit{high}-confidence bucket and drops to 21.1\% and 18.2\% in the \textit{medium} and \textit{low} buckets, respectively. These results suggest that developers may benefit from filtering out lower-confidence predictions to prioritize higher-quality results.

To quantify the relationship between confidence and actual FL success more precisely, we compute the Point-Biserial correlation between \name's continuous confidence scores and the binary success labels. The resulting correlation is 0.23 ($p<0.001$), which is weaker than the 0.57 reported in the original \orgname paper, yet still indicates a statistically significant positive association.
We further assess the calibration of \name's confidence score, i.e., its alignment with the true probability of FL success, using the Brier score~\cite{brier1950verification}, which quantifies the mean squared error between predicted confidence values and binary success labels (lower is better). \name yields a Brier score of 0.357, notably worse than a naive predictor that always outputs 0.5 (Brier score of 0.25), indicating that the raw confidence values are poorly calibrated.

To address this, following recent LLM calibration work~\cite{virk2025calibration}, we apply Platt Scaling~\cite{platt1999probabilistic}, a post-hoc calibration technique that fits a logistic regression model to map predicted probabilities (here, confidence scores) to observed outcomes. We perform 5-fold cross-validation to obtain rescaled confidence scores across all crash instances. This calibration reduces the Brier score to 0.199, demonstrating that rescaling can significantly improve the alignment between confidence and actual success, making the scores more interpretable and actionable for developers.

\subsection{RQ4. Contribution of Provided Functions}

\begin{figure}[t]
    \centering
    \includegraphics[width=0.95\linewidth]{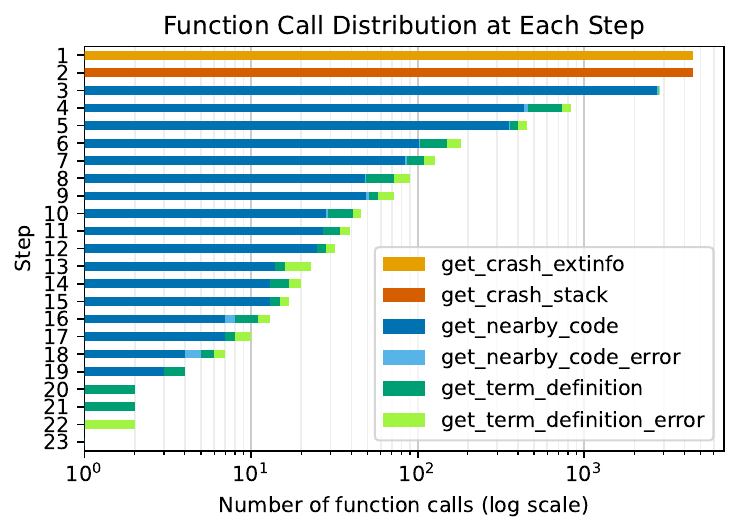}
    \caption{Distribution of function calls made by \name across reasoning steps.}
    \Description{This figure highlights how \name leverages different tools during the reasoning process.}
    \label{fig:RQ4_pattern}
\end{figure}

To understand \name's patterns of tool usage across all 4,540 \name runs (454 crashes $\times$ 10 repeated runs), we analyze which function the LLM invoked at each step of the reasoning process. On average, \name issued 3.1 function calls per run (maximum=23), indicating that it frequently terminates early (i.e., stops requesting further function calls and directly provides a root-cause explanation), far before exhausting its function-calling budget.
\Cref{fig:RQ4_pattern} shows the distribution of function usage across the reasoning steps over all \name runs. We observe that, as advised by the input prompt, \texttt{get\_crash\_extinfo} and \texttt{get\_crash\_stack} are invoked in the early steps, after which the LLM begins to examine additional information, such as the nearby code referenced in the stack trace. As intended, once it inspects the code, the LLM also attempts to locate the definitions of specific terms found within the inspected lines by calling \texttt{get\_term\_definition}, which in most cases (99.6\% in our evaluation) naturally triggers an additional call to \texttt{get\_nearby\_code}.

\begin{table}[t]
    \centering
    \caption{Distribution of \texttt{get\_term\_definition\_error} cases.}
    \scalebox{0.9}{
    \begin{tabular}{l|r}
    \toprule
    Error Type & Count \\
    \midrule
    Term does not appear in the given line & 263 \\
    File not found & 28 \\
    Term appears in the line, but navigation failed & 15 \\
    File found but invalid line reference & 3 \\
    Invalid argument format & 4 \\
    \midrule
    \textbf{Total} & \textbf{313} \\
    \bottomrule
    \end{tabular}}
    \label{tab:RQ4_term_errors}
\end{table}

\Cref{fig:RQ4_pattern} also shows that some function invocations failed. Specifically, \texttt{get\_nearby\_code} failed in 77 cases (2\%), and \texttt{get\_term\_definition} failed in 313 cases (40\%).  We further categorized the causes of these failures. All 77 \texttt{get\_nearby\_code} errors resulted from the LLM requesting code for files that could not be located. Among these, 26 and 21 cases involved absolute and relative paths that were not resolvable within the project context, respectively. The remaining 30 cases appeared to involve hallucinated paths, where the requested filenames were not found and not even present in the stack trace.
The 313 failures of \texttt{get\_term\_definition} are summarized in \Cref{tab:RQ4_term_errors}. In most cases, the LLM attempted to resolve the location of a symbol that did not exist at the specified code line. A contributing factor appears to be difficulty in interpreting the line-numbered code format produced by \texttt{get\_nearby\_code}, where line numbers are prepended to each line of the snippet. Additional failure cases included requests for non-existent files and language server being unable to resolve the symbol location.

\begin{table}[t]
    \centering
    \caption{Ablation results for \name under two configurations: disabling deep search by omitting \texttt{get\_term\_definition}, and using sanitized sections instead of raw contents.}
    \scalebox{0.9}{
    \begin{tabular}{c|r|rrrrr}
    \toprule
    Technique & Len. & \texttt{a@1} & \texttt{a@2} & \texttt{a@3} & \texttt{a@5} & \texttt{a@10}\\
    \midrule
    \multirow{2}{*}{\name}     & \multirow{2}{*}{3.83}  & 134 & 195 & 227 & 245 & 255 \\
                               & & (30\%) & (43\%) & (50\%) & (54\%) & (56\%)\\\midrule
    \name & \multirow{2}{*}{3.59} & 135 & 199 & 224 & 243 & 248 \\
     w/o Deep Search    &     & (30\%) & (44\%) & (49\%) & (54\%) & (55\%) \\\midrule
    \name & \multirow{2}{*}{3.88} & 130 & 189 & 212 & 244 & 250 \\
     w/ Sanitization   &      & (29\%) & (42\%) & (47\%) & (54\%) & (55\%)\\
    \bottomrule
    \end{tabular}
    }
    \label{tab:RQ4_ablation}
\end{table}

To more accurately assess the role of deep search, we disable \name's \texttt{get\_term\_definition} functionality and examine the resulting performance change. Interestingly, as shown in \Cref{tab:RQ4_ablation}, accuracy at lower $k$ values slightly improves in this configuration. This can be attributed to the previously observed fact that the LLM may struggle to effectively leverage the deep search tool in certain cases. Without the tool, the LLM may instead focus more narrowly on analyzing the stack trace itself, leading to improved precision for top-ranked predictions. However, performance at higher $k$ values declines when deep search is disabled. Upon further inspection, we found that with deep search enabled, \name included the correct buggy file in its FL rankings in 8 cases where the file's basename did not appear in the original stack trace. In these cases, the LLM identified relevant symbols in stack-trace contained files and successfully traced their definitions to external files, which were then marked as suspicious. Although these files were not ranked at the top, their inclusion demonstrates the potential value of deep search in uncovering non-obvious fault locations.
These findings indicate that while the current interface between the LLM and the deep search tool may not always be effective, further refinement, e.g., improving symbol resolution prompts or tool response formats, could enhance the utility of deep search and ultimately improve \name's overall performance.

As discussed in \Cref{sec:parser}, we also evaluate the effect of providing \name with sanitized crashdump content via \texttt{get\_crash\_stack} and \texttt{get\_crash\_extinfo}, instead of the raw section contents. This experiment is motivated by the observation that sanitization substantially reduces the number of input tokens, thereby lowering inference cost and improving compatibility with context length limits.
For instance, the raw \texttt{CRASH\_STACK} section contains an average of 9,326 tokens, which drops to 1,873 tokens after sanitization. Similarly, the \texttt{CRASH\_EXTINFO} section is reduced from 240 to 205 tokens. As shown in \Cref{tab:RQ4_ablation}, overall FL performance is slightly lower when using sanitized inputs, but the drop is negligible.
These results suggest a trade-off between cost and performance. Sanitized content may be preferable when resource constraints are a concern or when raw inputs exceed the model's context length. Indeed, among the 454 crash instances, we encountered one case where the raw \texttt{CRASH\_STACK} input triggered a context limit error, and sanitization was necessary for \name to function properly.

\section{Explanation Reliability Assessment}
\label{sec:root_cause_explanation}
For resolved crashes at \company, post-mortem root cause summaries are generated by LLMs. These summaries synthesize information from multiple failure-related artifacts, including the issue report (title, description, and comments), the ground-truth fix diff, and contents from relevant crashdump sections such as \texttt{BUILD}, \texttt{CRASH\_STACK}, and \texttt{CRASH\_SHORTINFO}. We refer to such summaries as \textit{postmortem analyses}.

Since \name produces root cause explanations before pinpointing the actual file location (as illustrated in \Cref{fig:diagram}), a natural question arises: \textit{Do the explanations generated by \name semantically align with the postmortem analysis?} If they do align, explanations from \name could provide significant insight to developers, as if they were hearing a description of the bug after it was resolved. To answer this, we generate a consolidated explanation for each file suggested by \name. More specifically, for every flagged file, we collect all explanations generated across runs where that file appears in the final prediction, and aggregate them into a single summary using \texttt{gpt-4o}.
For example, suppose we run \name three times for a given bug, and obtain predictions: $\{F_a\}$ (with explanation $E_1$), $\{F_a, F_b\}$ (with $E_2$), and $\{F_c\}$ (with $E_3$). In this case, the explanation for suspecting file $F_a$ is derived by summarizing $E_1$ and $E_2$.

To assess the semantic alignment between the aggregated explanations and the corresponding postmortem analyses, we forgo lexical similarity metrics (e.g., token overlap, BLEU), which fail to capture deeper semantic correspondence. Instead, we adopt an LLM-as-a-judge protocol~\cite{zheng2023llmasjudge}, leveraging the reasoning capabilities of LLMs to evaluate semantic alignment. For each crash file, we present the postmortem analysis alongside the \name-generated explanation summary and prompt the LLM (here, \texttt{gpt-4o}) to issue a categorical judgment: \texttt{aligned} or \texttt{misaligned}. To enhance robustness, we collect five independent judgments (each from a separate LLM invocation) and assign the final label based on majority vote. We report two alignment metrics. First, the overall alignment rate, defined as the proportion of crashes where at least one file-level explanation is judged as being aligned with the postmortem analysis, is 37\%. 
For example, the postmortem analysis for one \texttt{SIBABRT} crash was as follows:
\begin{quote}
\textit{“The crash was caused by a \textbf{race condition} in the job execution system (...). The issue arises when an exception occurs in one job while multiple jobs are being processed in parallel (...)”}
\end{quote}
While \name produced the following aligned explanation:
\begin{quote}
\textit{“\texttt{<BUGGY\_FILE>} could exhibit problematic behavior if the job queue management system does not properly synchronize access to job queues (...). Potential issues might include \textbf{race conditions} or (...).”}
\end{quote}
Second, acknowledging that developers are most likely to consult higher-ranked files, we report the top-1 and top-3 alignment rates, i.e., the fraction of crashes for which at least one explanation among the top 1 or top 3 ranked files is judged aligned with the postmortem analysis. This occurs in 113 out of 454 crashes for the top-1 explanation, yielding a top-1 alignment rate of 25\%. For the top-3 explanations, alignment is observed in 154 crashes, corresponding to a top-3 alignment rate of 34\%.

Overall, it is encouraging that \name could generate explanations that aligned with postmortem analyses -- which were generated with full information about the crash -- for a non-negligible portion of bugs. At the same time, it leaves significant room for improvement. While the LLM-as-a-judge procedure provided promising initial results, we note it is not clear what practitioners need from explanations; this knowledge would help improve the utility of explanations in future work.



\section{Conclusion}
\label{sec:conclusion}
In this work, we introduced \name, an LLM-driven FL approach that extends \orgname and is tailored for large-scale industrial crash debugging. Unlike traditional SBFL or MBFL techniques that rely on costly runtime data, \name operates directly on crashdumps and source code, making it practical in production environments where coverage and reproduction are often unavailable.
Our evaluation on 454 real-world \company crashes shows that \name consistently outperforms heuristic baselines across diverse crash types, with particularly strong results for SIGABRT and non-NPE segmentation faults. 
These findings highlight the potential of LLM-based agents as effective crash debugging assistants. At the same time, challenges remain in calibrating confidence scores and improving deep-search interactions. We see \name as a first step towards reliable integration of LLM-based FL into industrial crash-debugging workflows, and we hope this study inspires further research on LLM agents in realistic, information-constrained settings.


\balance
\bibliographystyle{ACM-Reference-Format}
\bibliography{refs}

\end{document}